\documentclass{elsart}

 \usepackage{graphicx}

\usepackage{amssymb}

\begin{document}

\begin{frontmatter}



\title{A solenoidal electron spectrometer for a precision measurement
of the neutron $\beta$-asymmetry with ultracold neutrons}
\author[caltech,kentucky]{B.\ Plaster,}
\author[caltech]{R.\ Carr,}
\author[caltech]{B.\ W.\ Filippone,}
\author[winnipeg]{D.\ Harrison,}
\author[caltech]{J.\ Hsiao,}
\author[caltech,lanl]{T.\ M.\ Ito,}
\author[caltech]{J.\ Liu,}
\author[caltech,winnipeg]{J.\ W.\ Martin,}
\author[caltech]{B.\ Tipton,}
\author[caltech]{J.\ Yuan}
\address[caltech]{W.\ K.\ Kellogg Radiation Laboratory, California
Institute of Technology, Pasadena, CA 91125, USA}
\address[kentucky]{Department of Physics and Astronomy, University of
Kentucky, \\ Lexington, KY 40506, USA}
\address[winnipeg]{Physics Department, University of Winnipeg,
Manitoba, Canada R3B 2E9}
\address[lanl]{Los Alamos National Laboratory, Los Alamos, NM 87545, USA}

\begin{abstract}
We describe an electron spectrometer designed for a precision
measurement of the neutron $\beta$-asymmetry with spin-polarized
ultracold neutrons.  The spectrometer consists of a 1.0-Tesla
solenoidal field with two identical multiwire proportional chamber and
plastic scintillator electron detector packages situated within
0.6-Tesla field-expansion regions.  Select results from performance
studies of the spectrometer with calibration sources are reported.
\end{abstract}

\end{frontmatter}

\section{Introduction}
\label{sec:introduction}

Nuclear $\beta$-decay studies have contributed significantly to our
understanding of the charged-current weak interaction and, more
generally, the fundamental symmetries underlying the Standard Model of
electroweak interactions.  Today, a rich program of $\beta$-decay
studies of nuclei, free neutrons, and pions continues in earnest, with
recent high-precision results providing important information on
Standard Model parameters and constraints on proposed
beyond-the-Standard-Model new physics scenarios
\cite{nico05,severijns06,abele08}.

In neutron $\beta$-decay, measurements of angular correlation
parameters and the lifetime continue to probe the limits of the
vector--axial-vector ($V-A$) description of the weak interaction.
Assuming the validity of the $V-A$ theory, the differential decay rate
distribution of the electron and neutrino momenta and the electron
energy in spin-polarized neutron $\beta$-decay (averaged over the
final-state electron spin) can be written in terms of angular
correlation parameters as \cite{jackson57}
\begin{equation}
\frac{dW}{dE_{e}d\Omega_{e}d\Omega_{\nu}}
\propto N(E_{e}) \left[1 + a \frac{\vec{p}_{e} \cdot \vec{p}_{\nu}}
{E_{e}E_{\nu}} + \langle\vec{\sigma}_{n}\rangle \cdot
\left( A\frac{\vec{p}_{e}}{E_{e}} + B\frac{\vec{p}_{\nu}}{E_{\nu}}
\right) \right],
\label{eq:distribution-correlations}
\end{equation}
where $E_{e}$ ($E_{\nu}$) and $\vec{p}_{e}$ ($\vec{p}_{\nu}$) denote,
respectively, the electron (neutrino) energy and momentum; $E_{0}$
($=782$ keV) denotes the electron endpoint energy; $N(E_{e}) =
p_{e}E_{e}(E_{0} - E_{e})^{2}$ is the electron energy spectrum;
$m_{e}$ is the electron mass; and $\langle \vec{\sigma}_{n} \rangle$
denotes the neutron polarization.  Neglecting recoil-order
corrections, the correlation parameters $a$ (the
$e$-$\overline{\nu}_{e}$-asymmetry), $A$ (the $\beta$-asymmetry), and
$B$ (the $\overline{\nu}_{e}$-asymmetry) are functions only of
$\lambda \equiv g_{A}/g_{V}$, defined as the ratio of the weak
axial-vector, $g_{A}$, and vector, $g_{V}$, couplings of the nucleon
to the $W$-boson \cite{wilkinson82,gardner01}.  Therefore,
measurements of $a$, $A$, and $B$ determine $\lambda$, by itself, a
fundamental parameter of the weak interaction.  In addition, a value
for $\lambda$, combined with a value for the lifetime, $\tau_{n}$,
where \cite{czarnecki04}
\begin{equation}
\frac{1}{\tau_{n}} = \frac{G_{\mu}^{2}|V_{ud}|^{2}}{2\pi^{3}}
m_{e}^{5}\left(1 + 3\lambda^{2}\right)\left(1 + \mathrm{RC}\right)f,
\label{eq:tau-lambda}
\end{equation}
permits an extraction of the CKM quark-mixing matrix element $V_{ud}$,
given a value for the Fermi constant $G_{\mu}$ as determined in muon
decay.  Here, $f$ is a phase-space factor, and RC represents the total
effect of all electroweak radiative corrections relative to muon decay
\cite{czarnecki04}.  A value for $V_{ud}$ derived solely from neutron
$\beta$-decay observables is particularly compelling for a first-row
CKM unitarity test \cite{abele04}, because the result is insensitive
to corrections for nuclear structure that must be applied to
extractions of $V_{ud}$ from measurements of $ft$ values in
superallowed $0^{+} \rightarrow 0^{+}$ nuclear $\beta$-decay
transitions \cite{hardy05}.

In the remainder of this paper, we describe a solenoidal electron
spectrometer designed to extract the neutron $\beta$-asymmetry
parameter $A$ from measurements of the angular correlation between the
neutron spin and the decay electron's momentum in the $\beta$-decay of
spin-polarized ultracold neutrons.  This experiment, the UCNA
experiment \cite{ucna} at the Los Alamos National Laboratory, will
yield the first-ever measurement of $A$ with ultracold neutrons (UCN),
offering different systematics than previous and ongoing measurements
employing beams of polarized cold neutrons \cite{abele02}.  The
ultimate goal of the UCNA experiment is a 0.2\% measurement of $A$.
We begin, in Section \ref{sec:requirements}, by detailing the
requirements on the spectrometer imposed by the consideration of
systematic uncertainties inherent to a measurement of $A$.  We follow
this in Section \ref{sec:design-construction} with a discussion of the
spectrometer design and construction.  Results from performance
studies of the spectrometer with conversion-electron sources are
presented in Section \ref{sec:calibration}.  We then conclude with a
brief summary in Section \ref{sec:summary}.

\section{Spectrometer requirements}
\label{sec:requirements}

\subsection{Overview of experiment}
\label{sec:requirements-overview}

\begin{figure}[t!]
\begin{center}
\includegraphics[angle=270,scale=0.45,clip=]{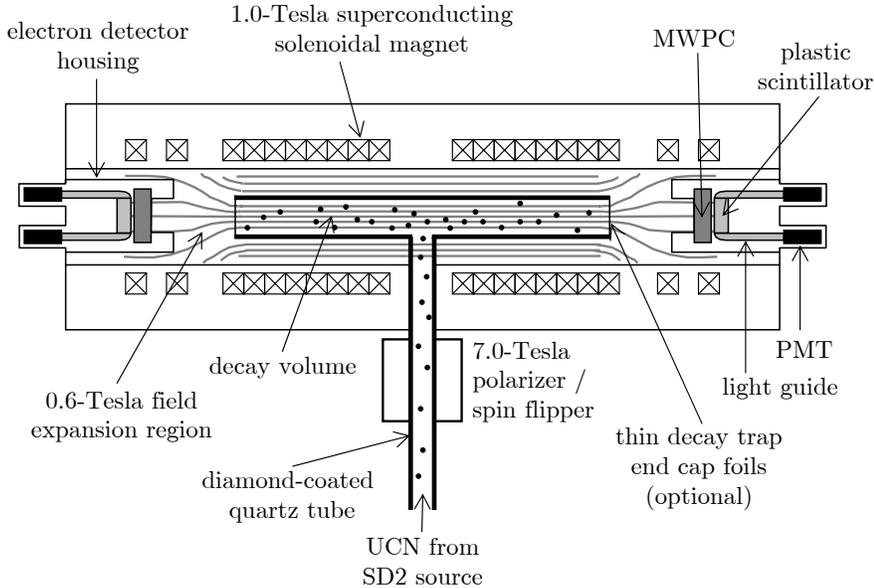}
\end{center}
\caption{Schematic diagram of the UCNA $\beta$-asymmetry experiment.}
\label{fig:ucna_schematic}
\end{figure}

To provide some context for a discussion of the spectrometer
requirements, a schematic diagram of the UCNA experiment is shown in
Fig.\ \ref{fig:ucna_schematic}.  UCN are produced in a superthermal
solid deuterium source \cite{morris02,saunders04}, spin-polarized via
transport through a 7-Tesla field, and then directed to the center of
a decay trap, where a fraction of them decay.  The decay trap consists
of a 3-m long, 10-cm diameter, diamond-coated quartz tube situated
within the warm bore of the solenoidal spectrometer.  Note that as an
option to increase the UCN density in the spectrometer (and, hence,
the $\beta$-decay rate), the ends of the decay trap can be closed off
with thin ($< 1$ $\mu$m) foils.  The spectrometer field strength of
1.0-Tesla is oriented along the axis of the decay trap (defining the
spin direction) and generated by a superconducting solenoidal magnet.
Two identical multiwire proportional chambers \cite{ito07} and plastic
scintillator detector packages are mounted on both ends of the
spectrometer and situated within 0.6-Tesla field-expansion regions.

Depending on their direction of emission, the decay electrons spiral
along the solenoidal field lines towards one of the two electron
detector packages, distributed over energy and angle according to
\begin{equation}
dW(\beta,\theta) \propto 1 + P_{n}A\beta\cos\theta,
\end{equation}
with $P_{n}$ the neutron polarization, $\beta$ the electron velocity
relative to $c$, and $\theta$ the angle of emission relative to the
neutron spin (or magnetic field).  The $\beta$-asymmetry $A$ is then,
in principle, extracted from the (energy-dependent)
experimental-asymmetry, $A_{\mathrm{exp}}(E_{e})$, measured in the two
detectors' count rates,
\begin{equation}
A_{\mathrm{exp}}(E_{e}) = \frac{N_{1}(E_{e})-N_{2}(E_{e})}
{N_{1}(E_{e})+N_{2}(E_{e})} = P_{n} A \beta \langle \cos \theta \rangle.
\label{eq:Aexp}
\end{equation}
The expected magnitude of the asymmetry is $A_{\mathrm{exp}} \sim
4.5$\% for $P_{n} = 1$, $\langle\cos\theta\rangle = 1/2$, and $\beta
\sim 0.75$ (for $E_{e} \sim 300$ keV, near the peak of the
$\beta$-decay spectrum).  In practice, ratios of the two detectors'
count rates for neutrons polarized parallel and anti-parallel to the
field [via transport through an adiabatic-fast-passage (AFP)
spin-flipper field region] are formed to extract the asymmetry via a
``super-ratio'' technique.  Here, the experimental asymmetry,
$A_{\mathrm{exp}}(E_{e})$, is extracted from the super ratio, $R$,
according to
\begin{eqnarray}
R &=& \frac{N_{1}^{\downarrow}(E_{e}) \cdot N_{2}^{\uparrow}(E_{e})}
{N_{1}^{\uparrow}(E_{e}) \cdot N_{2}^{\downarrow}(E_{e})}, \nonumber \\
A_{\mathrm{exp}}(E_{e}) &=& \frac{1 - \sqrt{R}}{1 + \sqrt{R}} =
P_{n} A \beta \langle \cos \theta \rangle,
\label{eq:super-ratio}
\end{eqnarray}
where $\uparrow$ and $\downarrow$ denote the two neutron spin states.
The advantage of the super-ratio technique is that differences between
the two detectors' efficiencies and variations in the UCN density for
the two spin states cancel in the ratio.

\subsection{Field uniformity requirement}
\label{sec:requirements-field-uniformity}

For a charged particle in a magnetic field $B$, the magnetic flux
enclosed by the particle's orbit (i.e., $Br_{L}^{2}$, with $r_{L}$ the
Larmor radius) is an adiabatic invariant~\cite{jackson}.  From this,
it can be shown that $p_{\perp}^{2}/B$, with $p_{\perp}$ the
transverse momentum, is also an adiabatic invariant.  The angle of
emission relative to the magnetic field (or spin) direction is, of
course, simply $\theta = \cos^{-1}(p_{\parallel}/p)$, with
$p_{\parallel}$ the longitudinal momentum, and $p =
\sqrt{p_{\perp}^{2} + p_{\parallel}^2}$ the total momentum.
Hereafter, we will frequently refer to $\theta$ as the pitch angle for
the trajectory.

Two considerations which follow from the adiabatic invariant influence
the requirement on the field uniformity.  First, electrons emitted
with momentum $p_{0} = \sqrt{p_{\perp,0}^2 + p_{\parallel,0}^2}$ in
some local field $B_{0}$ will be reflected from field regions $B$ if
$B > B_{\mathrm{crit}} \equiv (p_{0}^{2}/p_{\perp,0}^{2})B_{0}$, thus
contributing to a false asymmetry.  Second, by this same process,
electrons emitted at large pitch angles in the vicinity of a local
field minimum will be trapped.  These trapped electrons will scatter
from residual gas isotropically and dilute the extracted value of $A$.
A detailed analysis of both of these considerations showed that the
effect on the asymmetry is negligible for a field uniformity of $\pm 5
\times 10^{-4}$~\cite{ucna}.

\subsection{Fiducial volume definition}
\label{sec:requirements-fiducial}

Unlike previous measurements of $A$ utilizing flight paths of cold
neutron beams from research reactors, the UCNA experiment employs a
novel technique where the polarized UCN are transported to the center
of a material-walled decay trap situated within the warm bore of the
electron spectrometer.  Consequently, $\beta$-decay electrons
originating near the walls of the decay trap can scatter from the
walls prior to impacting the electron detector packages, leading to a
distortion in the measured energy spectrum.  Thus, one of the
requirements for the electron spectrometer is the ability to
reconstruct the transverse $(x,y)$ position of the $\beta$-decay event
with resolution sufficient to define a fiducial volume with a minimal
loss of counting statistics.

Reconstruction of the $(x,y)$ position is achieved via readout of the
induced charge on the two cathode planes of the multiwire proportional
chamber (MWPC), oriented at right angles relative to each other.
Complete details concerning the MWPC's design, construction, and
initial offline performance tests may be found in \cite{ito07}.  With
the maximum Larmor diameter for an endpoint-energy electron (with
$90^{\circ}$ pitch angle) of 7.76~mm in the 1-Tesla field, allowing
for a (conservative) 10\% safety factor yields a fiducial volume
radius of $(50 - 7.76 \times 1.10)$~mm $= 41.5$~mm (or, a 69\%
fiducial volume).  As discussed in detail later in this paper,
position resolutions of $\sim 2$~mm have been achieved.  This
resolution permits identification of the 8.5~mm radial cut annulus
with high confidence.

\subsection{Suppression of backscattering events}
\label{sec:requirements-backscattering-suppression}

Identification of electron events that undergo Coulomb backscattering
from any of the MWPC components or the plastic scintillator is
important for an extraction of the $\beta$-asymmetry $A$
\cite{martin06,schumann08}.  Particularly problematic are those events
termed ``missed backscattering events'', in which the electron
backscatters from one detector package with energy deposition
\textit{below} that detector's threshold.  Such unidentifiable
backscattering events contribute to a systematic shift in the
extracted value of the asymmetry, and can only be corrected for in
simulation.  Thus, the spectrometer includes the following features
designed to suppress the total and missed backscattering fractions.
First, the backscattering probability increases rapidly with the angle
away from normal incidence.  Large pitch-angles are eliminated with
the field-expansion ratio of 0.6, which maps pitch angles of
$90^{\circ}$ in the 1.0-Tesla region to $51^{\circ}$ in the 0.6-Tesla
region.\footnote{For the same reason, the transverse $(x,y)$ position
of the trajectory in the 1.0-Tesla region maps to that in the
0.6-Tesla region via a $1/\sqrt{0.6}$ factor.}  Second, the placement
of a low-energy-threshold MWPC in front of the plastic scintillator
detector facilitates the identification of low-energy-deposition
backscattering events which would otherwise not trigger the plastic
scintillator (e.g., due to dead-layer\footnote{The scintillator's
dead-layer was measured offline with a variable energy 0--130~keV
electron gun and found to be $3.0 \pm 0.3$ $\mu$m
\cite{junhua-thesis}.} and/or threshold effects).  Finally, for fixed
energy, the backscattering probability increases with atomic number
$Z$ \cite{tabata71}.  Thus, the MWPCs were constructed with low-$Z$
materials \cite{ito07}.

Note that results from backscattering studies carried out with
conversion-electron sources are not reported in this paper.  Instead,
a detailed companion paper comparing these data with Monte Carlo
studies is in preparation \cite{backscattering-paper}.

\subsection{Energy reconstruction}
\label{sec:requirements-energy}

A measurement of the total electron energy is required, as the
experimental asymmetry, $A_{\mathrm{exp}}(E_{e})$ in Eq.\
(\ref{eq:Aexp}), is proportional to $\beta = v/c$.  The chosen
detector, plastic scintillator, offers a low average atomic number
$Z$, important for the suppression of backscattering events, and fast
timing, defining the event trigger.  The primary requirements for the
scintillator detector are as follows.  First, the scintillator must be
sufficiently thick to provide a total energy measurement of the entire
neutron $\beta$-decay spectrum up to the endpoint $E_{0}$.  Second,
high energy resolution is required, demanding a high photoelectron
collection efficiency in the spectrometer's magnetic field
environment.  Third, a low energy threshold is required, again
demanding a high photoelectron yield, and also a low dark rate.
Fourth, the scintillator's fast timing response, coupled to fast
photomultiplier tubes, permits the identification and separation of
backscattering events triggering both scintillators.  Fifth, the size
of the scintillator is mechanically constrained by the size of the
solenoidal spectrometer's warm bore.  Finally, the scintillator
resides in the 0.6-Tesla field expansion region, requiring that the
light be transported along light guides to a lower field region ($\sim
300$ Gauss) for readout with the photomultiplier tubes.

\subsection{Suppression of backgrounds}
\label{sec:requirements-backgrounds}

One primary advantage of a neutron $\beta$-asymmetry measurement with
ultracold neutrons produced via a low-frequency (typically, 1/10 to
1/20 Hz) pulsed spallation-driven superthermal solid deuterium source
is the supression of (or, near absence of) beam-related backgrounds.
Signal-to-noise ratios in previous $A$ experiments with cold neutron
beams from research reactors were typically 7:1~\cite{abele02}, with
gamma-ray induced events constituting the primary background.  The
desire for a high signal-to-noise experiment was one motivation for
the introduction of the MWPC, as the requirement of a coincidence
between the MWPC and the scintillator leads to a significant reduction
in gamma-ray backgrounds.  Further, as discussed below, the
scintillator is coupled adiabatically to a series of light guides
which transport the light to four photomultiplier tubes.  Requiring at
least a two-fold photomultiplier tube coincidence further suppresses
noise from dark currents and low-energy gamma backgrounds.

\section{Spectrometer design, construction, and operation}
\label{sec:design-construction}

\subsection{1-Tesla superconducting solenoidal magnet}
\label{sec:design-construction-magnet}

\begin{figure}[t!]
\begin{center}
\includegraphics[angle=270,scale=0.50,clip=]{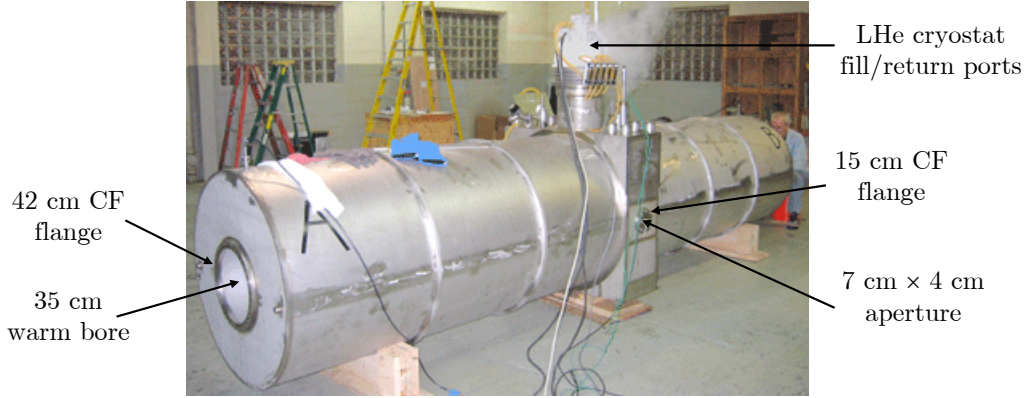}
\end{center}
\caption{(Color online) Photograph of the 1-Tesla superconducting
solenoidal magnet.  Provided courtesy of American Magnetics, Inc.}
\label{fig:scs_photo}
\end{figure}

A photograph of the as-built 1-Tesla superconducting solenoidal magnet
is shown in Fig.\ \ref{fig:scs_photo}.  In general terms, the magnet
is a warm-bore (35-cm diameter) 4.5-m long superconducting solenoid
with three 7-cm (vertical) $\times$ 4-cm (horizontal) rectangular
radial penetrations through the center of the coil.  One such
penetration is required for the transport of UCN to the decay trap (as
indicated schematically in Fig.\ \ref{fig:ucna_schematic}).  The
capacity of the magnet's liquid helium cryostat is $\sim 1600$ liters,
with a heat load during operation in persistence mode equivalent to a
boil-off rate of $\sim 10$ liters/hour.

The required radial penetration complicated the design of the
solenoid.  Analytic studies indicated that a single split-coil
solenoid would not satisfy the $\pm 5 \times 10^{-4}$ field
homogeneity requirement, but a multi-coil design with varied
currents/windings could be tuned to satisfy this requirement
\cite{junhua-thesis}.  The design proposed to the chosen vendor,
American Magnetics, Inc.\ (AMI) \cite{ami}, consisted of a
configuration of 32 50-cm diameter, 7-cm wide coils (each modeled as
70 current filaments spaced 1~mm apart) all separated by 7-cm, with
six different currents ranging from 1600--1800 A-turns
\cite{junhua-thesis}.  Ultimately, the designs of the coil
configurations for both the 1-Tesla region and the 0.6-Tesla
field-expansion region were proprietary.  The vendor's end-product
consisted of a main coil winding with a single persistence heater
switch, 28 shim coil windings (each with individual persistence heater
switches), and three rectangular 7-cm $\times$ 4-cm radial
penetrations (two providing horizontal access, one providing vertical
access, to the warm bore).

The cryostat for the magnet was fabricated by Meyer Tool and
Manufacturing, Inc.\ \cite{meyer}.  Although the cryostat was equipped
with a liquid nitrogen jacket buffer volume, in practice we have
forced the (colder) liquid helium boil-off through the liquid nitrogen
volume, as we have found this reduces the rate of liquid helium
boil-off.  Liquid helium is supplied by a helium liquefaction plant
operated solely for the UCNA experiment in a closed-loop mode.  This
plant consists of a Koch Model 1630 liquefier coupled to a 4000-liter
supply dewar, low-pressure ($\sim$ few psi) magnet boil-off return
lines coupled to 240-psi (supply-side) compressors, and a 240-psi
``medium pressure storage'' ambient temperature ballast tank with a
$\sim 2300$ liquid liters equivalent volume.

The power supply for the main coil, an AMI Model 12200PS-420
(200~A/12~V), functions as a slave to an AMI Model 420 Power Supply
Programmer unit.  The full energized field strength of 1.0 Tesla
requires a current of 124.0 A.  The power supply for the shim coils, a
bipolar Kepco Model BOP 20-10M ($\pm 20$ V/$\pm 10$ A), also functions
as a slave to a second AMI Model 420 Power Supply Programmer unit.
During magnet operations, the shim coils are energized one-by-one to
their optimized current values.  A LabView-based program installed on
a PC functions as the interface between the user and the Model 420
Power Supply Programmer units.

The MWPC/plastic scintillator electron detector packages, described in
detail below, are mounted to the two ends of the solenoidal magnet via
41.9-cm conflat flanges.  The central axis of the spectrometer is
located 2.74~m above the experimental floor.

\subsection{Multiwire proportional chambers}
\label{sec:design-construction-mwpc}

\begin{figure}[t!]
\begin{center}
\includegraphics[angle=270,scale=0.51,clip=]
{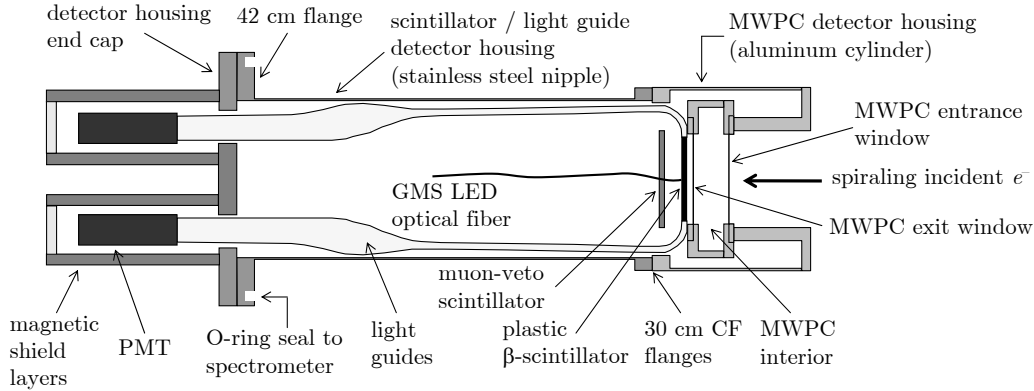}
\end{center}
\caption{Schematic diagram of the MWPC and plastic scintillator detector
package.}
\label{fig:detector_package_schematic}
\end{figure}

A schematic diagram illustrating the main components of the
MWPC/plastic scintillator detector package is shown in Fig.\
\ref{fig:detector_package_schematic}.  As noted already, the MWPCs are
relatively insensitive to gamma ray backgrounds, offer a low threshold
for the detection of backscattering events, and permit reconstruction
of the transverse $(x,y)$ coordinates of the $\beta$-decay events.
Although described in detail in \cite{ito07}, for completeness we
briefly review some of the MWPCs' most important features here.

First, to suppress missed backscattering events (again, those events
depositing no energy above threshold in the MWPC), the entrance window
separating the MWPC fill gas from the spectrometer vacuum was designed
to be as thin as possible.  Second, because of this thin entrance
window requirement, the fill gas pressure was required to be as low as
possible.  The chosen fill gas, C$_{5}$H$_{12}$ (2,2-Dimethylpropane,
or ``neopentane''), a low-$Z$ heavy hydrocarbon, was shown to yield
sufficient gain at a pressure of 100 Torr.  The minimum window
thickness for a 15-cm diameter window (equal to the diameter of the
plastic scintillator, discussed below) shown to support a 100 Torr
difference with a minimal leak rate (from pinholes) was 6 $\mu$m of
aluminized Mylar,\footnote{Note that when the spectrometer is
configured such that the ends of the decay trap are not closed off
with thin foils, the nominal 6 $\mu$m aluminized Mylar entrance window
is replaced with a UCN-reflective $^{58}$Ni-coated aluminized Mylar
window (to suppress contamination from neutron capture on $^{27}$Al).}
reinforced by Kevlar fiber.  Third, the MWPC mechanical size was
constrained by the requirement that the MWPC housing fit within the
solenoid's 35-cm warm bore diameter while also providing sensitivity
to the entire 10-cm diameter decay trap (projecting to a
$10/\sqrt{0.6} = 12.9$~cm diameter in the 0.6 Tesla field expansion
region where the MWPC/scintillator is mounted).  Finally, the anode
plane was strung with 64 10-$\mu$m diameter gold-plated tungsten
wires, and the cathode planes, separated from the anode plane by 10
mm, were strung with 64 50-$\mu$m diameter gold-plated aluminum wires.
The wire spacing on both the anode and the two cathode planes is 2.54
mm, yielding an active area of $16.3 \times 16.3$ cm$^{2}$.  This
active area maps to a $12.6 \times 12.6$ cm$^{2}$ square in the
1.0-Tesla region, providing full coverage of the 10-cm diameter decay
trap.

All of the anode wires are connected via conductor tracks on the anode
wireplane support frame; thus, the anode signal that is read-out is
the summation of the signals on the 64 individual wires.  With the
wires on one cathode plane oriented ``vertically'', providing
information on the ``horizontal'' coordinate (defined to be the
$x$-coordinate), and the wires on the other cathode plane oriented
``horizontally'', providing information on the ``vertical'' coordinate
($y$-coordinate), the $(x,y)$ position for each event is reconstructed
from the center-of-gravity of the two cathode planes' signals.

Note that the 64 wires on each of the two cathode planes are read-out
in groups of four.  This four-wire-grouping reduced the number of
electronics channels (thereby reducing the cost), but did not, as
shown later, degrade the resulting position resolution, as the typical
four-wire-group multiplicity is $\sim 3$--4 for the 368.1~keV
electrons from a $^{113}$Sn conversion-electron source.\footnote{The
368.1~keV energy we reference hereafter is the weighted mean (by
intensity) of the $^{113}$Sn K-, L-, and M-shell conversion-electron
energies of 363.8 keV (28.4\%), 387.5 keV (5.6\%), and 390.9 keV
(1.1\%).  The resolution of the scintillator is not sufficient to
distinguish the individual lines.}

\subsection{Plastic scintillator detectors}
\label{sec:design-construction-scintillator}

\subsubsection{Scintillator}
\label{sec:design-construction-scintillator-scintillator}

The plastic scintillator detector (or, hereafter, the
``$\beta$-scintillator'') is a 15-cm diameter, 3.5-mm thick disk of
Eljen Technology EJ-204 scintillator \cite{eljen}.  The EJ-204
wavelength of maximum emission is 408~nm.  This wavelength couples
well to a typical bialkili photocathode's wavelength sensitivity, and
is also sufficiently long for efficient optical transmission through a
UVT light guide.  Other desirable features of the EJ-204 scintillator
which motivated its choice include a high light output (68\% of
anthracene), long attenuation length (1.6~m), fast rise time (0.7~ns),
narrow pulse width (2.2~ns FWHM), and an index of refraction (1.58)
well-matched to that of UVT light guide (1.49).

The 15-cm diameter in the 0.6-Tesla field-expansion detector region
maps to a $15\sqrt{0.6} = 11.6$~cm diameter disc in the 1.0-Tesla
region of the spectrometer, providing full coverage of the 10-cm
diameter decay trap.  As discussed in~\cite{junhua-thesis}, the 3.5-mm
thickness was determined by two factors.  First, the range of an
$E_{0} = 782$~keV endpoint energy electron in plastic scintillator is
3.1~mm.  Thus, a total energy measurement requires a thickness of at
least 3.1~mm.  Second, the probability of a gamma ray interaction in
thin plastic scintillator is expected to scale with the scintillator
volume.  Thus, suppression of this background demands the thinnest
possible scintillator.\footnote{For the 3.5-mm thickness, the typical
energy deposition from cosmic-ray muons in the $\beta$-scintillator is
$\sim 700$--1500~keV (i.e., the peak is very broad).  Increasing the
scintillator thickness would have shifted the muon ``peak'' past the
neutron $\beta$-decay endpoint, but would have increased the total
background rate from gamma rays.  Note that muons are vetoed with a
scintillator located immediately behind the $\beta$-scintillator (see
Fig.\ \ref{fig:detector_package_schematic}).}  The chosen 3.5~mm
thickness included an additional thickness ``safety factor'' for the
energy measurement.

\subsubsection{Light guide system}
\label{sec:design-construction-scintillator-lightguide}

As indicated in Fig.\ \ref{fig:detector_package_schematic}, the light
from the scintillator disc is transported out of the 0.6-Tesla
field-expansion region over a distance of $\sim 1$~m along a series of
UVT light guides to photomultiplier tubes, situated in a region where
the magnetic field is $\sim 300$~Gauss.  Two options were considered
for light collection from the scintillator: (a) coupling light guides
directly to the scintillator back face, or (b) coupling light guides
to the scintillator edge.  Ultimately, edge collection was chosen for
the following reasons.  First, the surface area of the back face is a
factor $\sim 10$ greater than the surface area of the edge, requiring
an order-of-magnitude more photocathode coverage (leading to higher
cost, and also higher dark noise), while only yielding $\sim 18$\%
more light collection~\cite{junhua-thesis}.  Second, cosmic-ray muons
traversing the light guides generate \v{C}erenkov light.  Suppression
of this background favored the smaller light guide volume associated
with edge collection.

\begin{figure}[t!]
\begin{center}
\includegraphics[angle=270,scale=0.42,clip=]{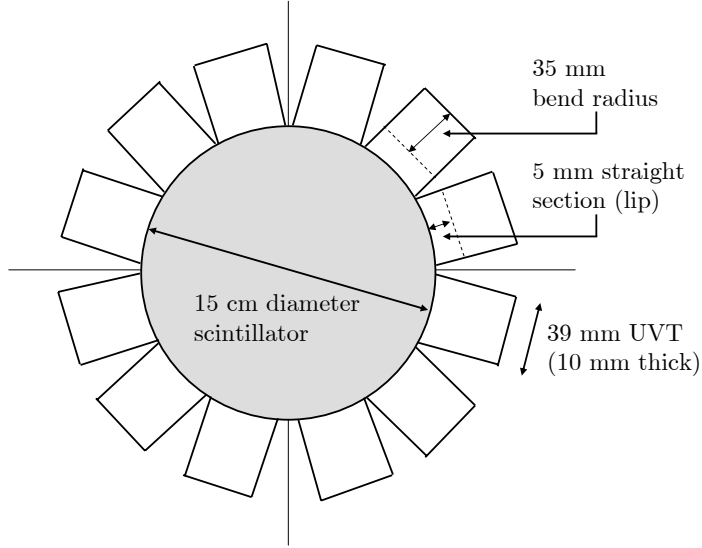}
\end{center}
\caption{Schematic diagram (front view) of the light guide system at
the scintillator edge interface.}
\label{fig:light_guide_schematic}
\end{figure}

\begin{figure}[t!]
\begin{center}
\includegraphics[angle=270,scale=0.50,clip=]{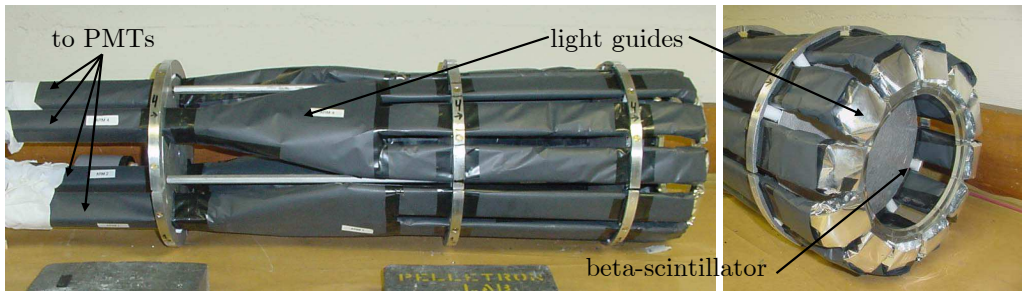}
\end{center}
\caption{(Color online) Photograph of the $\beta$-scintillator and
light guide assembly.}
\label{fig:light_guide_photo}
\end{figure}

The design for the edge collection light guide system was constrained
by the requirement that the detector system (shown in Fig.\
\ref{fig:detector_package_schematic}) fit within the spectrometer's
35-cm diameter warm bore, with an additional radial space allowance
for cabling and efficient vacuum pumping.  The design that was
implemented, shown schematically in Fig.\
\ref{fig:light_guide_schematic}, consisted of twelve rectangular
strips of 39-mm wide $\times$ 10-mm thick UVT light guide coupled to
the edge of the scintillator concentrically around its circumference
(via optical grease).  Each of these twelve strips includes a 5-mm
straight section, which couples the 3.5-mm thick scintillator edge to
the 10-mm thick light guide.  Following the 5-mm straight sections,
the light guides are then bent through $90^{\circ}$ over a 35-mm bend
radius and directed towards the photomultiplier tubes.  A novel
feature of the light guide transport system is that the twelve
rectangular strips adiabatically transform into four $39 \times
30$~mm$^{2}$ rectangular clusters, which are then coupled to the
(four) photomultiplier tubes.  The entire light guide system is
supported by an aluminum ring-like support frame.  A photograph of the
as-built light guide system, manufactured by Suzuno Giken
\cite{suzuno}, appears in Fig.\ \ref{fig:light_guide_photo}.

\subsubsection{Photomultiplier tubes}
\label{sec:design-construction-scintillator-pmts}

The choice of readout by four photomultiplier tubes (PMTs) was
motivated by the following considerations.  The effective light
collection surface area of the twelve rectangular light guides at the
scintillator edge interface is $12 \times (39 \times 10$~mm$^{2})$ =
4680~mm$^{2}$.  This surface area maps onto thirteen 1-inch PMTs,
three 2-inch PMTs, or one 3-inch PMT.  Readout by more than one PMT
was desired, as the requirement of a multi-PMT coincidence suppresses
dark noise and backgrounds from gamma-ray interactions in the
scintillator and \v{C}erenkov light from cosmic-ray muons traversing a
set of light guides coupled to the same PMT.  Ultimately, readout with
four 2-inch PMTs\footnote{Note that four (as opposed to three) 2-inch
PMTs are actually required in order to couple the rectangular geometry
of the light guides to the circular geometry of the photocathodes.}
was chosen as it provides for a multi-PMT coincidence, but requires
fewer electronics channels than readout with 1-inch PMTs.

The Burle 8850, a 12-stage photomultiplier with a bialkili
photocathode and a high-gain gallium-phosphide first dynode
\cite{burle-8850}, was chosen for the experiment.  This PMT was
favored as it is optimized for single-photon counting and features a
low dark noise rate, a fast rise time of 2.1~ns, a high gain of $1.6
\times 10^{7}$ at a nominal operating voltage of 2000~V (well below its
maximum operating voltage of 3000~V), and a peak quantum efficiency at
420~nm (well matched to the EJ-204 scintillator's 408~nm wavelength of
maximum emission).  The PMTs are glued to the ends of the rectangular
light guides.  High voltage for the PMTs is supplied with a LeCroy
HV4032 power supply unit.

\subsubsection{Photomultiplier tube magnetic shielding}
\label{sec:design-construction-scintillator-shielding}

\begin{figure}[t!]
\begin{center}
\includegraphics[angle=270,scale=0.50,clip=]
{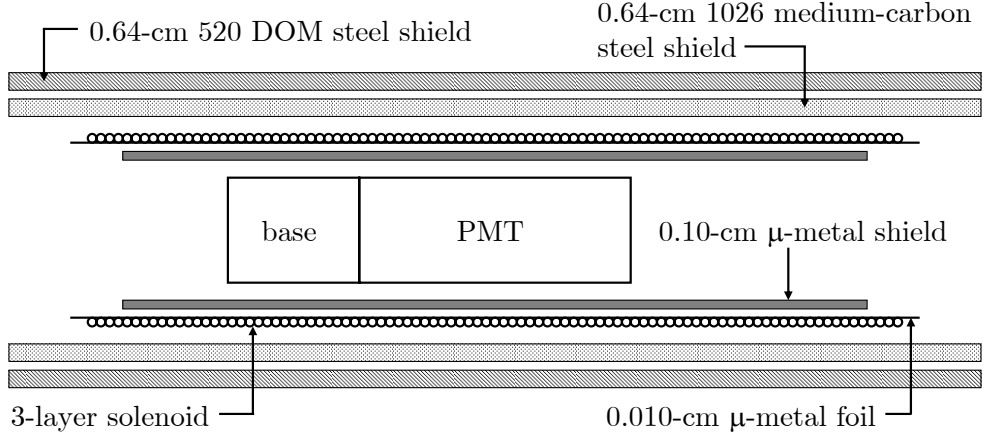}
\end{center}
\caption{Schematic diagram of the magnetic shielding for the
photomultiplier tubes.  See text for explanation of layers.}
\label{fig:magnetic_shielding_schematic}
\end{figure}

The magnetic field in the region of the PMTs is oriented primarily in
the axial direction, with a magnitude of $\sim 300$ Gauss.  This field
is shielded with an array of active and passive magnetic shielding
components, shown schematically in Fig.\
\ref{fig:magnetic_shielding_schematic}.  Moving from the exterior to
the interior, the first element is a 0.64-cm-thick 520 DOM steel
shield, with a nominal outer diameter of 11.4-cm.  The next element
is a close-fitting 0.64-cm-thick 1026 medium-carbon steel shield with
a nominal outer diameter of 10.2-cm.  The lengths of both of these
steel shields are 55.9-cm.  The third and fourth layers consist of a
$\sim 37.6$-cm-long 3-layer $\sim 325$-turn solenoid wound (with 18
AWG copper magnet wire) on the surface of a 7.6-cm outer diameter,
0.10-cm-thick cylinder of Advance Magnetics AD-MU-80 $\mu$-metal.
Note that because the $\mu$-metal shield is only 30.5-cm in length,
a 38.1-cm-long, 0.010-cm-thick sheet of AD-MU-80 $\mu$-metal foil was
wrapped on the surface of the 30.5-cm-long $\mu$-metal cylinder, with
the solenoids then actually wound on the surface of the thin
$\mu$-metal foil.

Bench tests of this magnetic shielding configuration with fields
comparable to the spectrometer fields in the region of the PMTs (i.e.,
up to $\sim 200$--400 Gauss) showed that residual shielded fields of
$\sim 0.1$ Gauss could be achieved with solenoid currents no greater
than 1.0~A.  Magnetic end caps were not required.

\subsection{Detector housing and spectrometer mount}
\label{sec:design-construction-mount}

The combined MWPC/plastic scintillator detector package is mounted
within a detector housing and inserted into the warm bore of the
1-Tesla solenoidal spectrometer.  After insertion, the
$\beta$-scintillator is located 2.2~m from the center of the UCN decay
trap.  As shown on the detector schematic in Fig.\
\ref{fig:detector_package_schematic}, the $\beta$-scintillator and the
light guide assembly are enclosed within a $\sim 64$-cm-long,
26.7-cm outer diameter stainless steel nipple with a 30.5-cm conflat
flange welded to one end, and a 41.9-cm flange (with an O-ring
groove) welded to the other end.  The MWPC is mounted to the nipple
via 30.5-cm conflat flanges.  The 41.9-cm flange on the other end of
the nipple is mounted to the 41.9-cm conflat flange on the
spectrometer warm bore.

Also shown on the schematic in Fig.\
\ref{fig:detector_package_schematic} are the magnetic shields for the
PMTs, which extend beyond the 41.9-cm flange on the end of the nipple.
The next-to-outermost shield layers, the 0.64-cm-thick 1026
medium-carbon steel shields, also serve as the vacuum seal (via O-ring
grooves) for the region beyond the 41.9-cm flange on the nipple.

\subsection{Gas flow and vacuum system}
\label{sec:design-construction-vacuum}

\begin{figure}[t!]
\begin{center}
\includegraphics[angle=270,scale=0.48,clip=]
{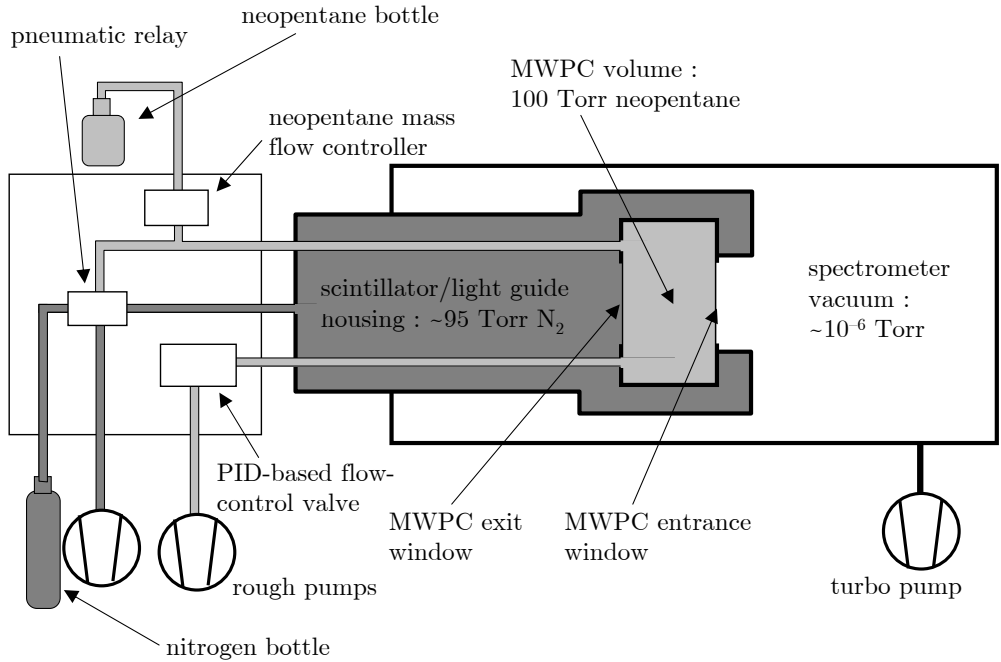}
\end{center}
\caption{Schematic diagram of the gas flow and vacuum system for the
MWPC and $\beta$-scintillator detector package coupled to the
spectrometer vacuum.}
\label{fig:vacuum_system_schematic}
\end{figure}

A schematic of the gas flow and vacuum system for the MWPC and
$\beta$-scintillator detector assembly is shown in Fig.\
\ref{fig:vacuum_system_schematic}.  The MWPC volume (nominally, 100
Torr of neopentane) is separated from the spectrometer vacuum by the
MWPC entrance window.  The detector housing for the
$\beta$-scintillator, light guide, and PMT assembly (nominally, $\sim
95$ Torr of nitrogen\footnote{As noted in \cite{ito07}, the pressure
of the nitrogen in the scintillator detector housing is maintained at
a somewhat lower pressure than the MWPC pressure.  This is to ensure
that the MWPC exit window bows out, or away, from the MWPC interior,
avoiding contact with the MWPC wire planes.})  is separated from the
MWPC volume by the MWPC exit window.  During normal operation, the
spectrometer, MWPC, and scintillator detector housing are first
evacuated at a rate of $< 1$~Torr/s, to minimize pressure gradients
across the MWPC windows.  Pressures in the spectrometer volume of
$\sim 10^{-6}$ Torr are typical.  The MWPC and scintillator detector
housing are then filled (at a rate significantly less than 1 Torr/s)
with neopentane and nitrogen, respectively, to their nominal operating
presures with the gas handling system described in \cite{ito07}.  Also
shown in \cite{ito07} is a photograph of the entire
MWPC/$\beta$-scintillator assembly on a rolling cart.

\subsection{Gain monitoring system}
\label{sec:design-construction-gms}

Possible temporal shifts in the $\beta$-scintillator's photomultiplier
tubes' gains are monitored, and subsequently corrected for, with a
gain monitoring system (GMS).  Separate (but identical) GMS systems
were constructed for the two $\beta$-scintillators.  Note that these
two separate systems are not redundant, as the gains of the PMTs in
the two detector packages can vary independently.  An overview of the
GMS system is as follows.

A Nichia Corporation UV LED (Model NSHU590) is enclosed within a
light-tight aluminum box.  The LED is driven in pulsed mode by a
voltage pulse from a CAEN Model C 529 (6-channel) LED driver installed
in a CAMAC crate, and controlled by a PC laptop.  The LED light is
transported along two fiber optic cables to the $\beta$-scintillator
(as indicated in Fig.\ \ref{fig:detector_package_schematic}) and to a
``monitor'' setup consisting of a NaI crystal viewed by a PMT.  A
gamma ray source ($^{60}$Co, 1.173 MeV and 1.333 MeV lines) is located
nearby.  The $^{60}$Co spectrum measured by the monitor PMT provides a
reference point for the stability of the amplitude of the LED output
pulse.  Variations in the LED spectra recorded in the
$\beta$-scintillator PMTs are then referenced to variations in the LED
spectrum relative to the $^{60}$Co spectrum observed in the monitor
PMT.

It can then be shown that the gain correction for some
$\beta$-scintillator PMT, $G_{\beta}(t)$, at some particular time $t$
relative to some starting time $t=0$ is
\begin{equation}
G_{\beta}(t) = \left(\frac{\mathrm{LED}_{m}(t)}{\mathrm{LED}_{m}(0)}\right)
\left(\frac{^{60}\mathrm{Co}_{m}(0)}{^{60}\mathrm{Co}_{m}(t)}\right)
\left(\frac{\mathrm{LED}_{\beta}(0)}{\mathrm{LED}_{\beta}(t)}\right).
\end{equation}
Here, $\mathrm{LED}_{m}(\ldots)$ denotes the response of the monitor
PMT to the LED pulse, $^{60}\mathrm{Co}_{m}(\ldots)$ the response of
the monitor PMT to the $^{60}$Co gamma ray spectrum, and
$\mathrm{LED}_{\beta}(\ldots)$ the response of the
$\beta$-scintillator PMT to the LED pulse.  The referencing of the LED
response to the $^{60}$Co response compensates for fluctuations in the
LED output.  Note that drifts in the gain of the monitor PMT are also
compensated, as the $^{60}$Co and LED spectra shift proportionally.

\subsection{Electronics and data acquisition}
\label{sec:design-construction-daq}

\subsubsection{Event trigger logic}
\label{sec:design-construction-daq-triggers}

The electronics for the experiment consists of a VME system for the
event trigger logic (via discriminators and programmable logic units)
and the readout of scalers, analog-to-digital convertors (ADCs), and
time-to-digital convertors (TDCs).  A NIM system coupled to the VME
system is employed for the implementation of a ``busy logic''.  This
busy logic vetos event triggers arriving during ADC and TDC conversion
times (i.e., while the ADC and TDC states are ``busy''), and was
implemented to prevent re-triggering (and subsequent readout of the
ADCs and TDCs) by scintillator afterpulses (single photoelectron
pulses from the slow-time component of the scintillator light
emission).  Bench studies showed that a veto time of $\sim 1$ $\mu$s
was sufficient to completely suppress contamination from re-triggering
by afterpulses \cite{junhua-thesis}.

\begin{figure}[t!]
\begin{center}
\includegraphics[angle=270,scale=0.52,clip=]
{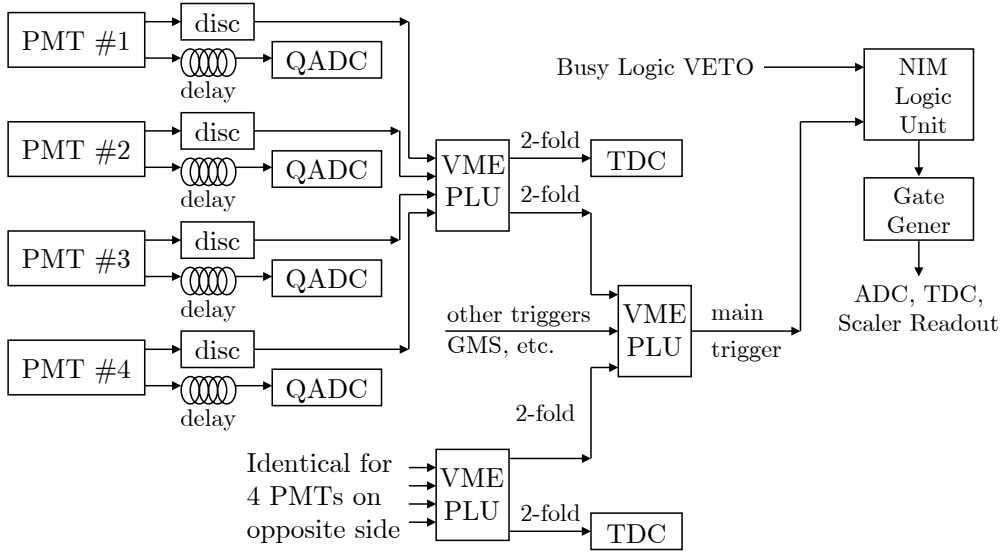}
\end{center}
\caption{Simplified schematic diagram of the trigger logic.  A
significant number of modules and circuit paths have been omitted.  As
noted in the text, the ``main trigger'' is an \texttt{OR} of all
possible event triggers.}
\label{fig:trigger_logic_schematic}
\end{figure}

A schematic diagram of the trigger logic is shown in Fig.\
\ref{fig:trigger_logic_schematic}, and an overview is as follows.  For
each detector package (detector 1 or detector 2), a trigger is defined
by a two-fold PMT coincidence requiring at least two PMTs with analog
signals above the discriminator threshold.  Note that the
discriminator threshold for each PMT is set at $\sim 1/2$ of the
amplitude of its single photoelectron peak.  An \texttt{OR} of a
detector 1 two-fold coincidence trigger, detector 2 two-fold
coincidence trigger, global GMS LED trigger, detector 1 GMS monitor
trigger (from $^{60}$Co events), detector 2 GMS monitor trigger
($^{60}$Co), and other experiment triggers then generates the main
event trigger.  The trigger logic for the two detectors' two-fold
coincidences and the main event trigger are performed with CAEN V495
Dual Programmable Logic Units (PLUs).  Those event triggers that are
not vetoed by the busy logic then trigger gate/delay generators for
the readout of the ADCs, TDCs, and scaler modules.

Among other variables, the TDCs (CAEN V775 modules with a 32-event
buffer memory, operated in \texttt{COMMON STOP} mode) provide a
relative measurement of the two detectors' two-fold coincidence
trigger arrival times.  Such relative timing information permits
identification of the earlier trigger (if two such triggers arrive),
important for the correct assignment of the initial direction of
incidence for backscattering events triggering both
$\beta$-scintillators.  Two types of ADC modules are used.  A
charge-integrating ADC (CAEN V792 QADC module with a 32-event buffer
memory), triggered for readout by a $\sim 140$~ns gate from a CAEN
V486 gate/delay generator, provides a measurement of the total charge
measured in each PMT.  The sum of the QADC spectra from the four PMTs
on each side is proportional to the energy deposition in the
scintillator.  Peak-sensing ADCs (CAEN V785 PADC modules with a
32-event buffer memory), triggered for readout by a $\sim 6$~$\mu$s
gate from a CAEN V462 gate generator, digitize the MWPC anode and
cathode-plane signals.  Scalers (CAEN V820 and V830 modules; the V830
module features a 32-event buffer memory) are used for deadtime
monitoring.

System deadtimes were measured offline as a function of the trigger
rate with a random pulser and found to be smaller than 10\% for event
rates up to $\sim 5$~kHz.\footnote{Beyond $\sim 5$~kHz, the data
acquisition system deadtime dominates the event loss.}  At lower rates
of $\sim 0.5$--1.0~kHz, the deadtime is proportional to the event
rate, of order $\sim 1$\%, and dominated by the ADC and TDC
digitization times.

\subsubsection{Data acquisition}
\label{sec:design-construction-daq-daq}

The data acquisition system is based on the MIDAS package
\cite{midas}.  Among many components, MIDAS features a buffer manager
for data flow management, a frontend acquisition code, a data logger
for the storage of data on disk, online analysis capabilities, a
web-browser-based run control, and an alarm system.  The UCNA
experiment utilizes a dedicated Linux-based workstation for
implementation of the frontend acquisition code, and another separate
Linux-based workstation for run control and online analysis.  The
frontend acquisition code accesses the VME system via a Struck PCI/VME
interface.\footnote{A Struck SIS1100 PCI card hosted by the Linux
workstation running the frontend acquisition code is connected via a
fiber optic link to a Struck SIS3100 VME board hosted by the VME
crate.}  The frontend continually polls the data-ready status of the
ADCs and TDCs.  Whenever any of the ADCs and/or TDCs have data ready
for readout in memory, the raw data from all ADC, TDC, and scaler
modules are transferred to the data buffer manager.  The MIDAS event
format structure consists of ``banks'' of raw data, which are
subsequently decoded by the MIDAS analyzer into formats readable by
the CERNLIB analysis packages ROOT and PAW (i.e., ``Ntuples'').

\section{Results from spectrometer performance studies}
\label{sec:calibration}

In this section we show select results from performance studies of the
electron spectrometer carried out with a source of $^{113}$Sn
368.1~keV conversion electrons with an event rate (above background)
as measured in the spectrometer of $\sim 85$~Hz.  The principal
analysis results shown below include energy spectra, position spectra,
an assessment of the spectrometer's position resolution, and an
assessment of the spectrometer's instrumental asymmetry.

\subsection{Solenoidal field map}
\label{sec:calibration-fieldmap}

\begin{figure}[t!]
\begin{center}
\includegraphics[scale=0.70]
{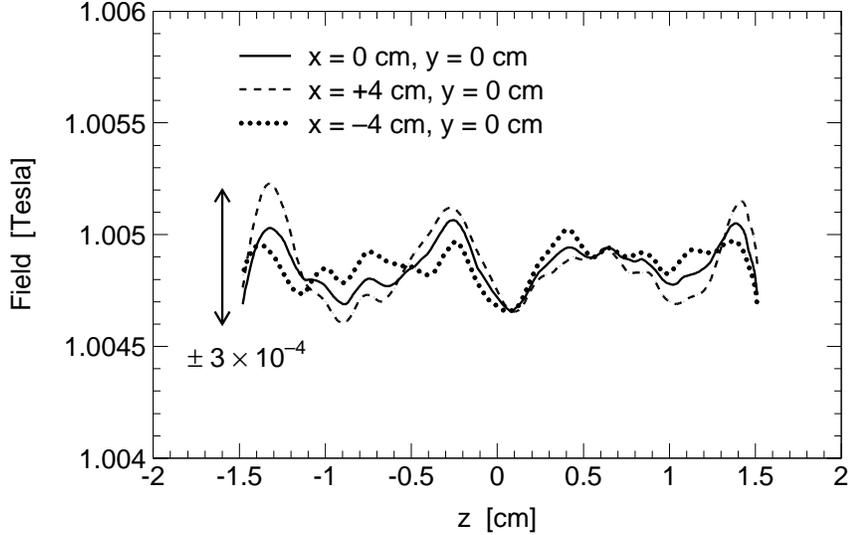}
\end{center}
\caption{Field maps from three different scans along the UCN decay
trap's 3-m length, showing uniformities at the level of $\pm 3 \times
10^{-4}$.}
\label{fig:scs_field_map_sept05}
\end{figure}

\begin{figure} [t!]
\begin{center}
\includegraphics[scale=0.70]
{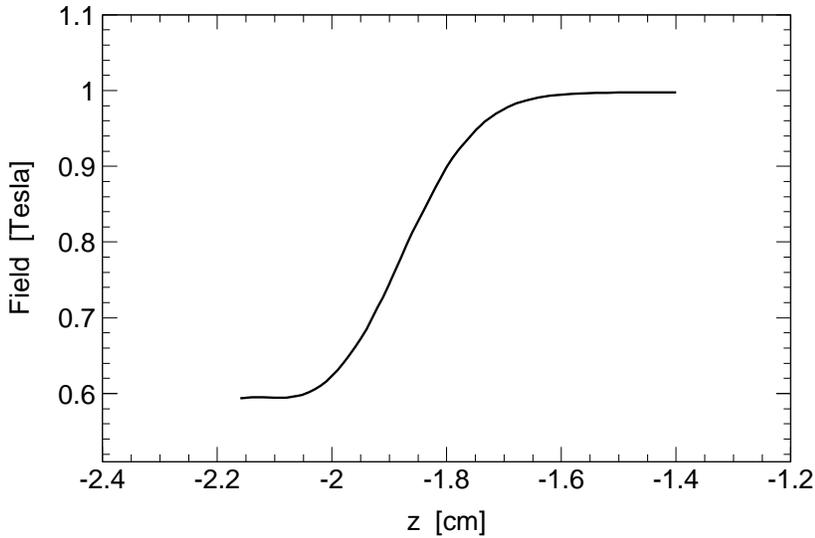}
\end{center}
\caption{Field map of the 1.0-Tesla to 0.6-Tesla field-expansion
region.}
\label{fig:field_expansion_map_sept05}
\end{figure}

First, however, we show results from field maps of the 1.0-Tesla
region (measured with an NMR probe) in Fig.\
\ref{fig:scs_field_map_sept05}.  As can be seen there, the field
profile satisfies the $\pm 5 \times 10^{-4}$ field uniformity
requirement over the 3-m extent of the UCN decay trap region.  Figure
\ref{fig:field_expansion_map_sept05} displays a field map of the
1.0-Tesla to 0.6-Tesla field-exansion region (measured with a Hall
probe).

\subsection{Energy spectra}
\label{sec:calibration-energyspectra}

\begin{figure}[t!]
\begin{center}
\includegraphics[scale=0.70]{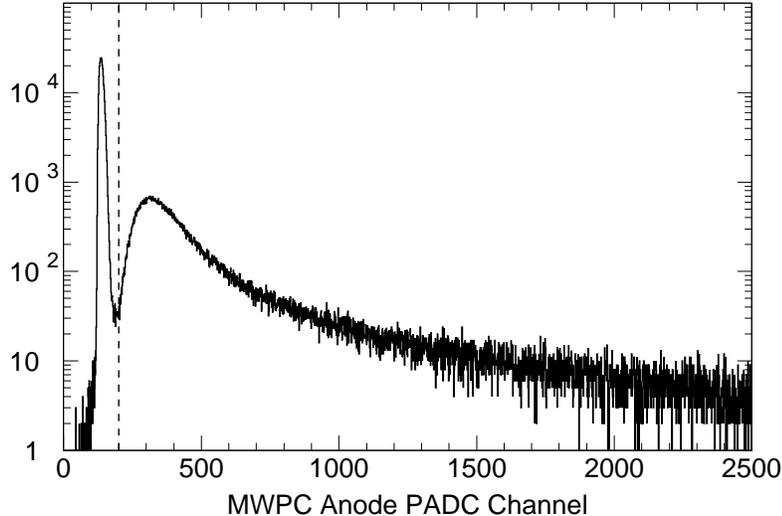}
\end{center}
\caption{Typical MWPC anode spectrum (PADC channel number) from a run
with the $^{113}$Sn source positioned within the fiducial volume.  The
dashed line indicates the typical position of the cut.}
\label{fig:west_anode_spectrum}
\end{figure}

\begin{figure}[h!]
\begin{center}
\includegraphics[scale=0.70]{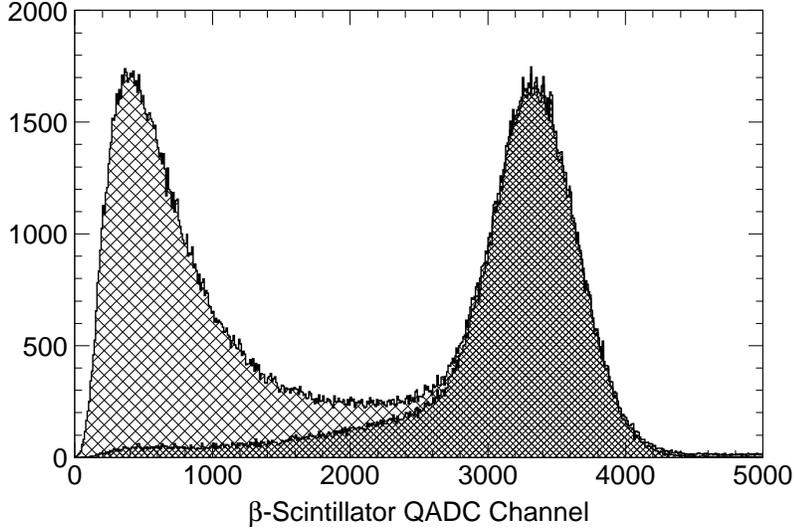}
\end{center}
\caption{Typical $\beta$-scintillator spectra from a run with the
$^{113}$Sn source positioned within the fiducial volume.  The
light-shaded (dark-shaded) histogram shows the spectrum before (after)
the cut on the MWPC anode response.}
\label{fig:west_scintillator_spectrum}
\end{figure}

A typical MWPC anode spectrum extracted from a run taken with the
$^{113}$Sn source positioned with the spectrometer's fiducial volume
is shown in Fig.\ \ref{fig:west_anode_spectrum}.  The
$\beta$-scintillator spectrum (summed over the response of the four
photomultiplier tubes) for this same run, shown before and after
application of the cut on the MWPC anode response, is shown in Fig.\
\ref{fig:west_scintillator_spectrum}.  As can be seen there, the
requirement of a coincidence between the MWPC and the
$\beta$-scintillator greatly reduces the background from gamma-ray
interactions.  The low-energy tail on the $^{113}$Sn spectrum is
primarily the result of energy loss in the MWPC entrance and exit
windows.\footnote{Note that for the data shown here, the MWPC entrance
and exit window thicknesses were 25 $\mu$m Mylar (non-aluminized), not
the optimized 6 $\mu$m thickness.}  Backscattering from the
MWPC/$\beta$-scintillator also contributes to the tail, but the
contribution of this process to the structure of the tail is small (as
the total detected backscattering event fraction is small, $\sim
4.6$\% for the 368.1~keV $^{113}$Sn conversion electron energies)
compared to the energy loss in the MWPC windows.

\begin{figure}[t!]
\begin{center}
\includegraphics[angle=270,scale=0.46,clip=]{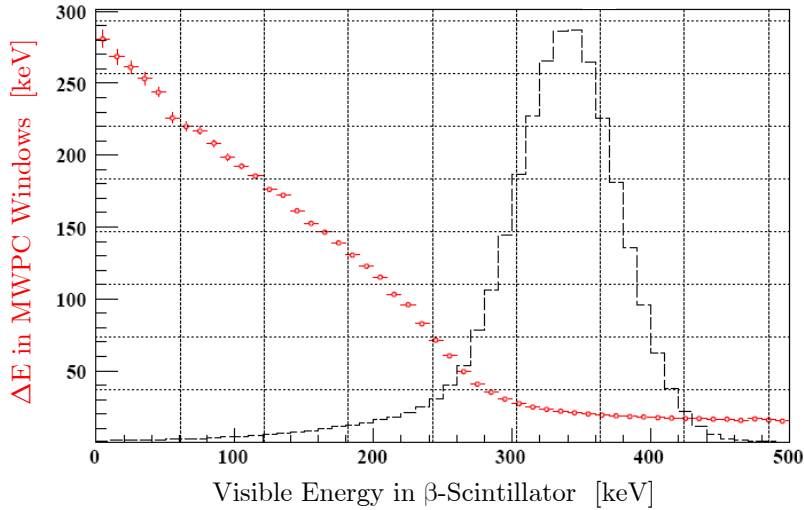}
\end{center}
\caption{\texttt{GEANT4} simulated results for energy loss from
$^{113}$Sn conversion electrons in the MWPC windows plotted versus the
visible energy deposition in the $\beta$-scintillator, overlaid with
a histogram of the visible energy distribution.}
\label{fig:sn_energy_tail}
\end{figure}

The impact of energy loss in the MWPC windows is illustrated in Fig.\
\ref{fig:sn_energy_tail}.  There, results from a \texttt{GEANT4}
simulation of the energy loss in the MWPC windows are plotted versus
the simulated visible energy\footnote{The visible energy is defined to
be the total energy loss in the $\beta$-scintillator active region
(beyond the dead layer) for those events generating a two-fold PMT
coincidence trigger.} deposition in the $\beta$-scintillator, and
overlaid with a histogram of the visible energy deposition.  Although
not shown there, the energy loss in the MWPC windows is correlated
with the initial pitch angle in the 1.0-Tesla region.

\subsection{Event types}
\label{sec:calibration-eventtypes}

\subsubsection{Definitions}
\label{sec:calibration-eventtypes-definitions}

\begin{figure}[t!]
\begin{center}
\includegraphics[angle=270,scale=0.58,clip=]{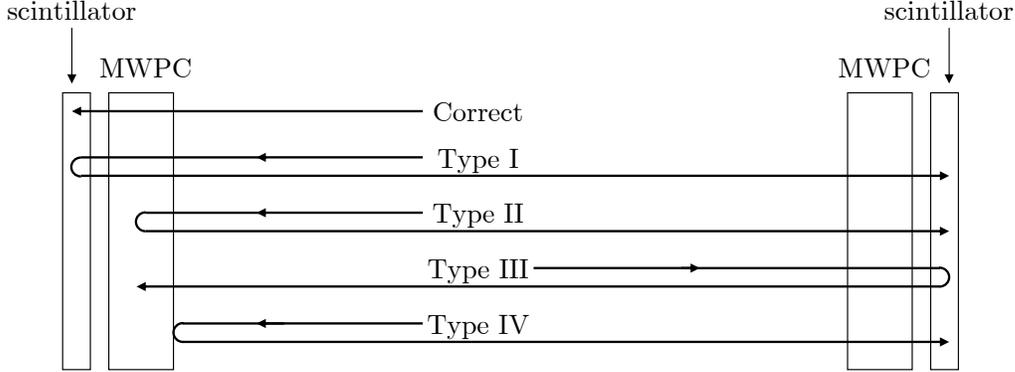}
\end{center}
\caption{Schematic diagram of the five event types.  See text for
descriptions.}
\label{fig:event_types}
\end{figure}

The $\beta$-asymmetry measurement requires an accurate determination
of the electron's initial direction of incidence.  Complicating the
extraction of the asymmetry are the five different event types, shown
schematically in Fig.\ \ref{fig:event_types}.

\begin{itemize}

\item
\textit{Correct events}: Events in which an electron, incident
initially on one of the $\beta$-detector packages, does not
backscatter from any element of that side's $\beta$-detector package,
and is then detected in that side's $\beta$-scintillator.

\item
\textit{Type I backscattering events}: Events in which an electron,
incident initially on one of the $\beta$-detector packages, deposits
sufficient energy to generate a two-fold trigger in that side's
$\beta$-scintillator, backscatters from that side's
$\beta$-scintillator, and is then detected in the opposite side's
$\beta$-scintillator.

\item
\textit{Type II backscattering events}: Events in which an electron,
incident intially on one of the $\beta$-detector packages, deposits
energy above the cut in that side's MWPC, backscatters either from
that side's MWPC (i.e., gas, wire planes, or back window) or
$\beta$-scintillator without depositing sufficient energy to generate
a two-fold trigger (e.g., backscatters from the scintillator's dead
layer), and is then detected in the opposite side's
$\beta$-scintillator.

\item
\textit{Type III backscattering events}: Events in which an electron,
incident initially on one of the $\beta$-detector packages, deposits
sufficient energy to generate a two-fold trigger in that side's
$\beta$-scintillator, backscatters from that side's
$\beta$-scintillator, deposits energy above the cut in the opposite
side's MWPC, and is then stopped in either the opposite-side MWPC's
gas, wire planes, or back window or the opposite-side's
$\beta$-scintillator's dead layer.

\item
\textit{Type IV backscattering events}: Events in which an electron,
incident initially on one of the $\beta$-detector packages,
backscatters from that side's MWPC without depositing energy above
the cut (e.g., backscatters from the MWPC entrance window), and is
then detected in the opposite side's $\beta$-scintillator.

\end{itemize}

Note that Type I backscattering events can be identified via
examination of the two detectors' two-fold coincidence trigger arrival
times, thus determining the initial direction of incidence
unambiguously (as discussed earlier in Section
\ref{sec:design-construction-daq-triggers}).  In contrast, simulation
input (regarding the energy deposition in the MWPCs) is needed to
distinguish Type II from Type III backscattering events.  For example,
without using any such simulation input (i.e., using only the
$\beta$-scintillator trigger to determine the direction of incidence),
the Type II and Type III events shown in Fig.\ \ref{fig:event_types}
would appear identical, with the direction of incidence for the Type
II event misidentified.  Type IV events cannot, of course, be
distinguished from Correct events; these are the ``missed
backscattering events'' discussed earlier in Section
\ref{sec:requirements-backscattering-suppression} which can only be
corrected for in simulation.

The analysis results presented hereafter are only for those events
identified by the event reconstruction as Correct events.  Because
this event sample also includes (unidentifiable) Type IV
backscattering events, we label these events ``Type 0 events''.  As
noted earlier, a detailed companion paper discussing the
identification of, and energy spectra of, the various backscattering
event types is in preparation \cite{backscattering-paper}.

\subsubsection{Selection of Type 0 events}
\label{sec:calibration-eventtypes-type 0}

\begin{figure}[t!]
\begin{center}
\includegraphics[scale=0.70]{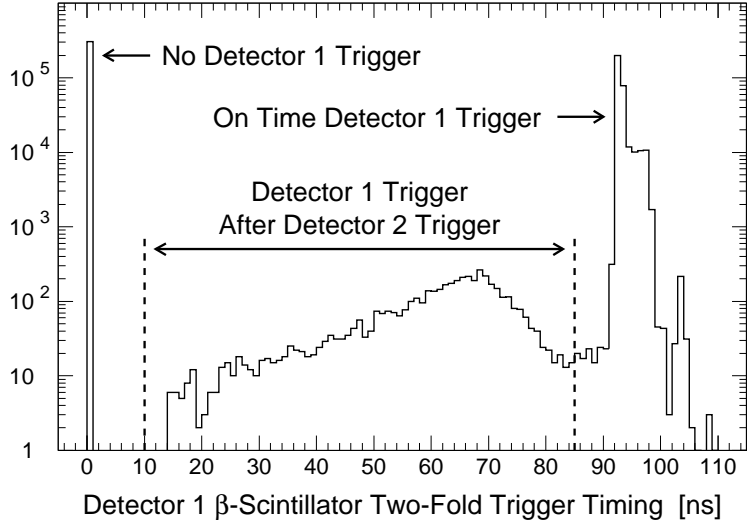}
\end{center}
\caption{Typical two-fold coincidence trigger arrival timing
spectrum.  See text for interpretation of the peaks.}
\label{fig:east_two_fold_timing}
\end{figure}

The trigger logic (shown in Fig.\ \ref{fig:trigger_logic_schematic})
is such that each two-fold trigger generates the \texttt{START} for
its respective TDC channel.  The main event trigger (as noted earlier,
an \texttt{OR} of the two-fold triggers and any other event triggers)
is delayed by $\sim 92$~ns and generates the \texttt{COMMON STOP} for
the TDC.  A typical two-fold coincidence arrival time spectrum (shown
here for detector 1 two-fold triggers) at the main trigger logic unit
is shown in Fig.\ \ref{fig:east_two_fold_timing}.  Main event triggers
generated by the arrival of a detector 1 two-fold trigger appear at
the delay time of $\sim 92$~ns (i.e., an ``on time'' trigger) ,
whereas main event triggers generated by the arrival of a detector 2
two-fold trigger, with no later arrival of a detector 1 two-fold
trigger, appear at the peak at 0~ns (i.e., a ``time-out'' for that TDC
channel).  Those events located between these two extremes (i.e., the
broad secondary peak located at $\sim 68$~ns) correspond to main event
triggers generated by the arrival of a detector 2 two-fold trigger at
the logic unit, with the later arrival of a detector 1 two-fold
trigger.  In other words, these are events which backscattered from
the detector 2 package into the detector 1 backage (i.e., Type I
backscattering events incident initially on the detector 2 package).
Note that a 150 keV electron will traverse the 4.4~m separation
between the two $\beta$-scintillators in 23.2~ns; thus, the $\sim
24$~ns separation between these two peaks is as expected.  Note that
the spectrum is cut off below $\sim 15$~ns because the dynamic range
of the TDC was set at its minimum value of 140~ns for this run.

Type 0 events were identified according to the following selection rules:

\begin{itemize}

\item
The initial incidence detector package's two-fold coincidence trigger
arrival time must reconstruct to the $\sim 92$~ns delay time.  The
other detector package's two-fold coincidence trigger arrival time
must reconstruct to a TDC ``time-out''.

\item
The MWPC on the $\beta$-scintillator hit side must display an anode
signal above the cut (as indicated in Fig.\
\ref{fig:west_anode_spectrum}).  The MWPC on the other side must
display an anode signal below the cut.

\item
The muon-veto scintillator (as shown in Fig.\
\ref{fig:detector_package_schematic}) on the $\beta$-scintillator hit
side must not record a hit.

\end{itemize}

\subsection{Position spectra}
\label{sec:calibration-positionspectra}

\subsubsection{Cathode-plane multiplicities}
\label{sec:calibration-positionspectra-multiplicities}

\begin{figure}[t!]
\begin{center}
\includegraphics[scale=0.70]{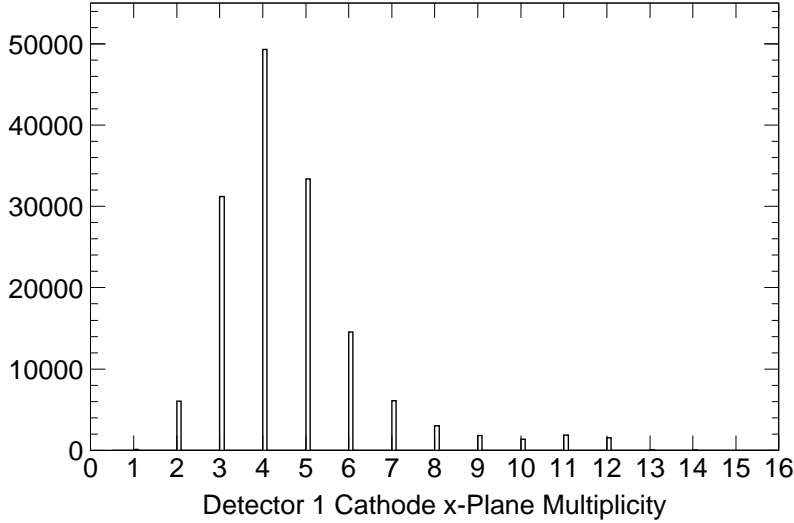}
\end{center}
\caption{Typical cathode-plane multiplicity (shown here for the
detector 1 $x$-plane) from a run with the $^{113}$Sn source.}
\label{fig:cathode_plane_multiplicity}
\end{figure}

As noted earlier, the 64 wires (with 2.54-mm spacing) on each of the
two MWPC cathode planes are read-out in groups of four.  Henceforth,
we will refer to each of these four-wire configurations simply as a
``wire''.  Despite this four-wire grouping, most of the events exhibit
a multiplicity (on each cathode plane) of $\geq 3$, where we define
the event-by-event cathode multiplicity to be the number of wires with
a signal above a prescribed threshold.  A typical distribution of
multiplicities from a run with the $^{113}$Sn source is shown in Fig.\
\ref{fig:cathode_plane_multiplicity}.  Here, the threshold was defined
by a two-sigma limit ($\sim 50$ channels) of the cathode pedestal
peaks, as determined by Gaussian peak fitting.

\subsubsection{Position reconstruction results}
\label{sec:calibration-positionspectra-algorithm}

The transverse $(x,y)$ positions are computed from the cathode-plane
wire spectra according to weighted sums,
\begin{equation}
x_{\mathrm{MWPC}} = \frac{\sum_{i} Q_{i} x_{i}}{\sum_{i} Q_{i}},~~~~~
y_{\mathrm{MWPC}} = \frac{\sum_{j} Q_{j} y_{j}}{\sum_{j} Q_{j}},
\end{equation}
where $Q_{i}$ denotes the signal above pedestal for the
$i^{\mathrm{th}}$ channel, $x_{i}$ and $y_{i}$ denote the $x$- and
$y$-coordinate of the $i^{\mathrm{th}}$ channel, and the sums $i$ and
$j$ run over all of the $x$- and $y$-plane cathode channels,
respectively, satisfying the two-sigma pedestal threshold.

\begin{figure}[t!]
\begin{center}
\includegraphics[angle=270,scale=0.52,clip=]
{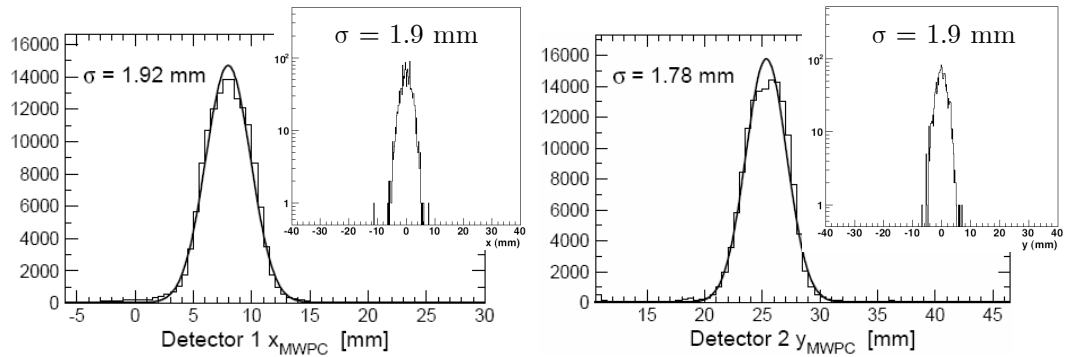}
\end{center}
\caption{Typical MWPC reconstructed position spectra for Type 0 events
from a run with the $^{113}$Sn source located within the fiducial
volume.  The solid curves are the results of Gaussian fits to the
peaks, with the fitted widths indicated.  Distributions obtained from
a \texttt{GEANT4} simulation of the MWPC response are shown in the
insets.}
\label{fig:mwpc_position_spectra}
\end{figure}

Sample MWPC reconstructed position spectra (shown here for detector 1
$x$-reconstruction and detector 2 $y$-reconstruction) for Type 0
events, extracted from a run with the $^{113}$Sn source located within
the fiducial volume, are shown in Fig.\
\ref{fig:mwpc_position_spectra}.  Note that the position spectra shown
here, and all spectra shown hereafter, have been projected back to the
spectrometer's 1.0-Tesla region.  Gaussian fits to the peaks are
plotted, showing fitted widths slightly better than $\sim 2$~mm.  The
widths of these distributions agree well with those extracted from a
\texttt{GEANT4} Monte Carlo simulation of the MWPC response to the
368.1~keV $^{113}$Sn conversion electrons (shown in the insets).

\begin{figure}[t!]
\begin{center}
\includegraphics[scale=0.43]{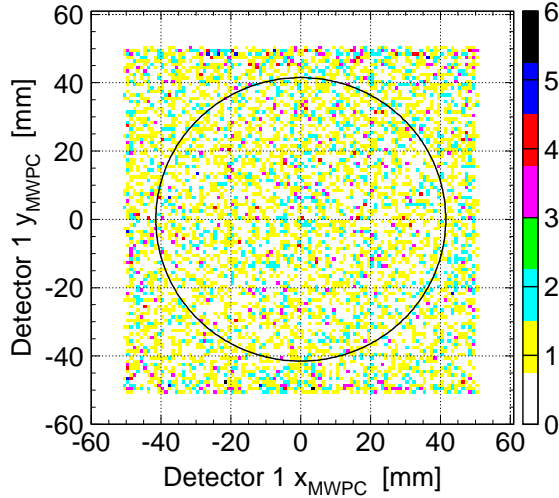}
\end{center}
\caption{(Color online) Two-dimensional plot of detector 1 $(x,y)$
positions for events tagged as muons by the muon-veto scintillator.
The circle denotes the extent of the 41.5-mm radius fiducial volume.}
\label{fig:muon_position_spectra}
\end{figure}

A two-dimensional plot of detector 1 $(x,y)$ positions are shown in
Fig.\ \ref{fig:muon_position_spectra} for those events identified by
the muon-veto scintillator as muons.  As expected, these background
events are distributed uniformly over the MWPC's sensitive area.

\subsection{Spectrometer position resolution}
\label{sec:calibration-resolution}

\begin{figure}
\begin{center}
\includegraphics[angle=270,scale=0.60]
{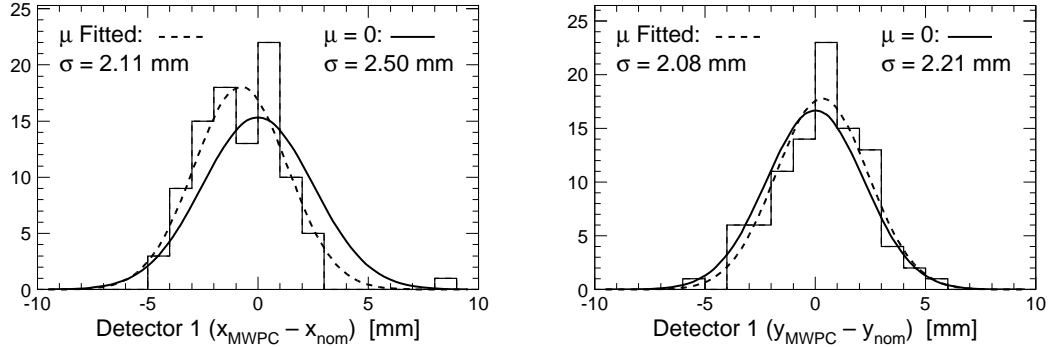}
\end{center}
\caption{Histograms of $(x_{\mathrm{MWPC}}-x_{\mathrm{nom}})$ and
$(y_{\mathrm{MWPC}}-y_{\mathrm{nom}})$ difference distributions for
reconstruction by the dectector 1 MWPC.  The dashed (solid) curves are
the results of Gaussian fits in which the means were fitted as a free
parameter (constrained to be 0).}
\label{fig:spectrometer_position_resolution}
\end{figure}

To assess the position response of the spectrometer, we analyzed data
from runs taken with the $^{113}$Sn source located at 96 different
positions within the fiducial volume.  We extracted the position
resolution of the spectrometer from distributions of differences
between the nominal (known) positions, and the positions reconstructed
by the MWPC [i.e., histograms of
$(x_{\mathrm{MWPC}}-x_{\mathrm{nom}})$ and
$(y_{\mathrm{MWPC}}-y_{\mathrm{nom}})$].  These difference
distributions are shown for the detector 1 reconstruction in Fig.\
\ref{fig:spectrometer_position_resolution}.

The dashed curves are the results of Gaussian fits to the
distributions, with the means fitted as free parameters.  Note that a
slight mispositioning ($\sim$ few mm) of the detectors relative to the
magnetic field lines may contribute to the non-zero fitted means.
Because we were interested in assessing the resolution relative to the
nominal positions, we also performed constrained-Gaussian fits to
these distributions, in which the means were fixed to be 0; the
results of these fits are shown as the solid curves.  Because the
differences between the fitted widths from the free- and
constrained-Gaussian fits were not great, we defined the average of
the two fitted widths to be the spectrometer position resolution.  The
resulting values are listed in Table
\ref{tab:spectrometer_position_resolution}.  These data have
demonstrated that the position resolution of the spectrometer is $\sim
2.0$~mm over the fiducial volume.

\begin{table}[t!]
\begin{center}
\caption{Extracted values of the spectrometer position resolution from
MWPC position reconstruction.  Listed are the fitted widths for the
free- and constrained-Gaussian fits to the reconstructed--nominal
position difference distributions.  The position resolution for each
coordinate is defined to be the average of the free and constrained
fitted widths.}
\begin{tabular}{lccc} \hline\hline
& \multicolumn{2}{c}{Fitted Widths $\sigma$ [mm]}& Position \\
Reconstruction& Free-Fit $\mu$& Constrained $\mu = 0$&
  Resolution [mm] \\ \hline\hline
Detector 1 MWPC $x$& 2.11& 2.50& 2.31 \\
Detector 1 MWPC $y$& 2.08& 2.21& 2.15 \\ \hline
Detector 2 MWPC $x$& 1.57& 2.14& 1.86 \\
Detector 2 MWPC $y$& 1.55& 1.45& 1.50 \\ \hline\hline
\multicolumn{4}{c}{Average Position Resolution = 1.96 mm} \\ \hline\hline
\end{tabular}
\label{tab:spectrometer_position_resolution}
\end{center}
\end{table}

\subsection{Spectrometer instrumental asymmetry}
\label{sec:calibration-asymmetry}

We analyzed the data from all of our $^{113}$Sn calibration runs and
calculated an instrumental asymmetry, denoted $\xi$, for Type 0 events
according to
\begin{equation}
\xi = \frac{N_{1}-N_{2}}
{N_{1}+N_{2}}.
\end{equation}
We found that the mean instrumental asymmetry for those runs
satisfying the fiducial volume radius cut of 41.5~mm is $\xi = (-0.400
\pm 0.083)$\%.

The instrumental asymmetry is itself $\sim 10$\% of the expected
physics asymmetry, $A_{\mathrm{exp}} =
P_{n}A\langle\beta\cos\theta\rangle \sim 4.5$\%.  The origin of a
non-zero instrumental asymmetry (from an isotropic source) is likely
the choice of cuts (e.g., cuts on the MWPC anode spectra, or cuts for
muon identification) for the event selection.  Although non-zero, the
instrumental asymmetry cancels to all orders in the extraction of the
$\beta$-asymmetry $A$ from the experimental super-ratio,
Eq.~(\ref{eq:super-ratio}), as the super-ratio asymmetry is insensitive to
differences between the two detectors' efficiencies.

\section{Summary}
\label{sec:summary}

In this paper we described a solenoidal electron spectrometer designed
for a precision measurement of the neutron $\beta$-asymmetry with
ultracold neutrons.  This spectrometer consists of a 1.0-Tesla
superconducting solenoid, with a combined MWPC and plastic
scintillator detector package for detection of the $\beta$-decay
electrons.  The UCNA experiment is currently operating at the
Los Alamos National
Laboratory, and collection of $\beta$-asymmetry data is ongoing.
Further experiments to perform a precise measurement of the neutron
$\beta$-decay energy spectrum, and to extract small energy-dependent
terms in the $\beta$-asymmetry due to recoil-order corrections, are
possible if the spectrometer's energy resolution can be improved.
This will be achieved by implementing large-area silicon detectors as
an alternate technology to the plastic scintillator detectors.  Other
experiments aimed at the determination of other angular correlation
parameters, such as $a$ and $B$ in Eq.\
(\ref{eq:distribution-correlations}), can be conducted if proton
detection can be incorporated into the apparatus.  We intend to
achieve proton detection by introducing thin
secondary-electron-emitting foils into the apparatus, into which the
decay protons would be accelerated by an electrostatic potential.
Both initiatives in detector technology development are ongoing at
this time.

We conclude this paper by emphasizing that our techniques and results
for the rejection of gamma-ray induced events and an assessment of the
spectrometer's position resolution are generally applicable to a
variety of precision $\beta$-decay experiments utilizing neutrons and
nuclei. \\

\textbf{Acknowledgments} \\ \\
We thank R.\ Cortez and J.\ Pendlay for their skillful technical
contributions to the design, fabrication, and deployment of the
detector systems.  We thank S.~Currie for his devoted efforts to the
operation and maintenance of the helium liquefaction plant.  We thank
the entire UCNA collaboration for many valuable suggestions.  This
work was supported in part by the National Science Foundation under
grant numbers PHY-0079767 (a Major Research Instrumentation Program
grant), PHY-0244899, PHY-0555674, and also by the Natural Sciences and
Engineering Research Council of Canada.




\begin{thebibliography}{99}





\bibitem{nico05} J.\ S.\ Nico and W.\ M.\ Snow, Annu.\ Rev.\ Nucl.\
         Part.\ Sci.\ \textbf{55}, 27 (2005).

\bibitem{severijns06} N.\ Severijns and M.\ Beck,
         Rev.\ Mod.\ Phys.\ \textbf{78}, 991 (2006).

\bibitem{abele08} H.\ Abele,
         Prog.\ Part.\ Nucl.\ Phys.\ \textbf{60}, 1 (2008).

\bibitem{jackson57} J.\ D.\ Jackson, S.\ B.\ Treiman, and H.\ W.\ Wyld, Jr.,
         Phys.\ Rev.\ \textbf{106}, 517 (1957).

\bibitem{wilkinson82} D.\ H.\ Wilkinson,
         Nucl.\ Phys.\ \textbf{A377}, 474 (1982).

\bibitem{gardner01} S.\ Gardner and C.\ Zhang,
         Phys.\ Rev.\ Lett.\ \textbf{86}, 5666 (2001).

\bibitem{czarnecki04}
         A.\ Czarnecki, W.\ J.\ Marciano, and A.\ Sirlin,
         Phys.\ Rev.\ D \textbf{70}, 093006 (2004);
         W.\ J.\ Marciano and A.\ Sirlin,
         Phys.\ Rev.\ Lett.\ \textbf{96}, 032002 (2006).

\bibitem{abele04} H.\ Abele \textit{et al}.,
         Eur.\ Phys.\ J.\ C \textbf{33}, 1 (2004).

\bibitem{hardy05} J.\ C.\ Hardy and I.\ S.\ Towner,
         Phys.\ Rev.\ Lett.\ \textbf{94}, 092502 (2005);
         J.\ C.\ Hardy and I.\ S.\ Towner,
         Phys.\ Rev.\ C \textbf{71}, 055501 (2005).

\bibitem{ucna} The UCNA experiment, A.\ Saunders and A.\ R.\ Young,
         spokespersons; R.~Carr \textit{et al}.,
         ``Technical review report for an accurate measurement of the
         neutron spin~-- electron angular correlation in polarized
         neutron beta decay with ultracold neutrons'' (2000).

\bibitem{abele02} H.\ Abele \textit{et al}.,
         Phys.\ Rev.\ Lett.\ \textbf{88}, 211801 (2002).

\bibitem{ito07} T.\ M.\ Ito \textit{et al}.,
         Nucl.\ Instrum.\ Methods Phys.\ Res.\ A \textbf{571}, 676 (2007).

\bibitem{morris02} C.\ L.\ Morris \textit{et al}.,
         Phys.\ Rev.\ Lett.\ \textbf{89}, 272501 (2002).

\bibitem{saunders04} A.\ Saunders \textit{et al}.,
         Phys.\ Lett.\ B \textbf{593}, 55 (2004).

\bibitem{jackson} J.\ D.\ Jackson, \textit{Classical Electrodynamics},
Third Edition (John Wiley \& Sons, Inc., New York, 1999).

\bibitem{martin06} J.\ W.\ Martin \textit{et al}.,
         Phys.\ Rev.\ C \textbf{68}, 055503 (2003);
         J.\ W.\ Martin \textit{et al}.,
         Phys.\ Rev.\ C \textbf{73}, 015501 (2006).

\bibitem{schumann08} M.\ Schumann and H.\ Abele,
         Nucl.\ Instrum.\ Methods Phys.\ Res.\ A \textbf{585}, 88 (2008).

\bibitem{tabata71} T.\ Tabata, R.\ Ito, and S.\ Okabe,
         Nucl.\ Instrum.\ Methods \textbf{94}, 509 (1971).

\bibitem{backscattering-paper} Manuscript in preparation.

\bibitem{junhua-thesis} J.\ Yuan, Ph.D.\ thesis, California Institute
         of Technology (2006).

\bibitem{ami} American Magnetics, Inc.,
         \texttt{www.americanmagnetics.com}.

\bibitem{meyer} Meyer Tool and Manufacturing, Inc.,
         \texttt{www.mtm-inc.com}.

\bibitem{eljen} Eljen Technology, \texttt{www.eljentechnology.com}.

\bibitem{suzuno} Suzuno Giken,
         \texttt{http://www.apace-science.com/tie\_up/suzuno.htm}.

\bibitem{burle-8850} Burle Technologies, Inc., \\
         \texttt{www.burle.com/cgi-bin/byteserver.pl/pdf/8850.pdf}.

\bibitem{midas} MIDAS Data Acquisition System,
         \texttt{midas.psi.ch}~~or~~\texttt{midas.triumf.ca}.

\end{thebibliography}
\end{document}